\documentclass[rmp,twocolumn]{revtex4}
\usepackage{amssymb,epsf}
\begin{document}
\title
{ Classification of scale free networks\\}

\author
{ K.-I. Goh$^*$, E.S. Oh$^*$, H. Jeong$^{\dag}$, B. Kahng$^*$, 
and D. Kim$^*$\\}

\affiliation
{
$^*$School of Physics and Center for Theoretical Physics, 
Seoul National University, Seoul 151-747, Korea\\
$^{\dag}$Department of Physics, Korea Advanced Institute of Science and 
Technology, Daejon 305-701, Korea\\}

\begin{abstract}
While the emergence of a power law degree distribution in complex 
networks is intriguing, the degree exponent is not universal. 
Here we show that the betweenness centrality displays a power-law 
distribution with an exponent $\eta$ which is robust and use it to 
classify the scale-free networks. 
We have observed two 
universality classes with $\eta \approx 2.2(1)$ and $2.0$, 
respectively. 
Real world networks for the former are the protein 
interaction networks, the metabolic networks for eukaryotes and 
bacteria, and the co-authorship network, and those for the 
latter one are the Internet, the world-wide web, and the metabolic 
networks for archaea. 
Distinct features of the mass-distance relation, generic topology
of geodesics and resilience under attack of the two classes are identified.
Various model networks also belong to either of the two classes
while their degree exponents are tunable.\\
\end{abstract}

\maketitle

Emergence of a power law in the degree distribution 
$P_D(k)\sim k^{-\gamma}$ in complex networks is an interesting 
self-organized phenomenon in complex systems \cite{strogatz,review1,review2}. 
Here, the degree 
$k$ means the number of edges incident upon a given vertex.
Such a network is called scale-free (SF) \cite{physica}. 
Real world networks which are SF include 
the author collaboration network \cite{newman} in social systems, 
the protein interaction network (PIN) \cite{pin}
and the metabolic network \cite{metabolic} in biological systems,
and the Internet \cite{internet} and the world-wide web (WWW) \cite{www,ha} in communication systems.
The power-law behavior means that most vertices 
are sparsely connected, while a few vertices are intensively connected 
to many others and play an important role in functionality. 
While the emergence of such a SF behavior in degree 
distribution itself is surprising, the degree exponent $\gamma$ 
is not universal and depends on the detail of network structure. 
As listed in Table 1, numerical values of the exponent $\gamma$ 
for various systems are diverse but most of them are in the range 
$2 < \gamma \le 3$. From the viewpoint of theoretical physics, 
it would be interesting to search a universal quantity associated with
SF networks. 

Recently a physical quantity called ``load" was introduced as a 
candidate for the universal quantity in SF networks.
It quantifies the load of a vertex in the transport of data packet 
along the shortest 
pathways in SF networks \cite{load}. 
It was shown that the load distribution exhibits a power law, 
$P_L(\ell)\sim \ell^{-\delta}$, and the exponent $\delta$ is 
robust as $\delta \approx 2.2$ for diverse SF networks 
with various degree exponents in the range $2 < \gamma \le 3$.
Since the universal behavior of the load exponent was obtained 
empirically, fundamental questions such as how the load exponent 
is robust in association with network topology or the 
possibility of any other universal classes existing, 
have not been explored yet. In this paper, 
we address those issues in detail.

While the load is a dynamic quantity,
it is closely related to a static quantity, the ``betweenness 
centrality (BC)", commonly used in sociology to quantify
how much a given person is influential in a society \cite{betweenness}. 
To be specific, BC is defined as follows. Let us consider the set of 
the shortest pathways, or geodesics, between a pair of vertices $(i,j)$ 
and denote their number by $C(i,j)$.
Among them, the number of the shortest pathways running through 
a vertex $k$ is denoted by $C_k(i,j)$. 
The fraction $g_k(i,j)=C_k(i,j)/C(i,j)$ may be interpreted as 
the amount of the role played by the vertex $k$ in social relation between 
two persons $i$ and $j$. Then the BC of the vertex $k$ is defined 
as the accumulated sum of $g_k(i,j)$ over all ordered pairs for 
which a geodesic exists, {\it i.e.}, 
\begin{equation}
g_k=\sum_{i\neq j}g_k(i,j)=\sum_{i\neq j}\frac{C_k(i,j)}{C(i,j)}.
\end{equation}
Because of only slight difference between load and BC, 
both quantities behave very closely.
In fact, the BC $g_k$ of each vertex is exactly the same as the
load for tree graphs. In general, distributions of the two 
are indistinguishable within available resolutions.
The BC distribution follows 
a power law, 
\begin{equation}
P_B(g) \sim g^{-\eta}, 
\end{equation}
where $g$ means BC and the exponent $\eta$ is the same as the load 
exponent $\delta$.
Since the topological feature of a network can be grasped more 
transparently using BC, we deal with BC in this work. 

Based on numerical measurements of the BC exponent for a variety 
of SF networks, we find that SF networks can be classified into 
only two classes, say, class I and II. For the class I, the BC 
exponent is $\eta \approx 2.2(1)$
and for the class II, it is $\eta\approx 2.0(1)$.
We conjecture the BC exponent for the class II 
to be exactly $\eta=2$ since it can be derived analytically
for simple models.
We show that such different universal behaviors in the BC 
distribution originate from different generic topological 
features of networks. 
Moreover, we study a physical problem, the resilience 
of networks under an attack, showing different behaviors 
for each class, as a result of such difference in generic topologies.
It is found that the networks of class II are much more vulnerable 
to the attack than those for class I.

To obtain our results, we use available data for real world networks
or existing algorithms for model networks.
Once a SF network is constructed, we select a pair of vertices 
$(i,j)$ on the network and identify the shortest pathways between 
them. Next, BC is measured on each vertex along the shortest 
pathways using the modified version of the breath-first search 
algorithm introduced by Newman \cite{newmanalgorithm}.
We measure the BC distribution and the exponent $\eta$ for a 
variety of networks both in real world and in silico.\\

\noindent
{\sf
Real World and Artificial Networks Investigated\\}
The networks that we find to belong to the class I with $\eta=2.2(1)$ 
include:
(i) The co-authorship network in the field of the neuroscience, 
published in the period 1991-1998 \cite{coauthor}, where 
vertices represent scientists and they are connected if 
they wrote a paper together. 
(ii) The protein interaction network of the yeast {\it S. cerevisiae} 
compiled by Jeong {\it et al.} \cite{pin} (PIN1), 
where vertices represent proteins and the two proteins
are connected if they interact\footnote{The network 
is composed of disconnected clusters of different 
sizes, {\it viz.}, small isolated clusters as well as a giant cluster. 
For both (ii) and (xi), the degree 
distribution is likely to follow a power law but there needs an 
exponential cutoff to describe its tail behavior for finite 
system. However, it converges to a clean power law for (xi) 
as system size increases, but the converging rate is rather slow
 \cite{jeenu}. 
Despite this abnormal behavior in the degree distribution 
for finite system, the BC distribution follows a pure power law 
with the exponent $\eta \approx 2.2(1)$ in (ii) and (xi). 
}.
(iii) The core of protein interaction network of the yeast {\it S. cerevisiae} 
obtained by Ito {\it et al.}\footnote{In contrast to (ii), the degree distribution obeys a power law.}
(PIN2) \cite{ito}.
(iv) The metabolic networks for 5 species of eukaryotes and 
32 species of bacteria in Ref.\cite{metabolic}, where 
vertices represent substrates and they are connected if a
reaction occurs between two substrates via enzymes.
The reaction normally occurs in one direction, so that  
the network is directed. 
(v) The Barab\'asi-Albert (BA) model \cite{ba} when the number 
of incident edges of an incoming vertex $m \ge 2$.  
(vi) The geometric growth model by Huberman and Adamic \cite{ha}. 
(vii) The copying model \cite{copying}, whose degree exponent is in the range of $2 < \gamma \le 3$. 
(viii) The undirected or the directed static model \cite{load}, whose degree exponent is in the range of $2 < \gamma \le 3$ or $2 < (\gamma_{\rm in}, \gamma_{\rm out}) \le 3$, respectively.  
(ix) The accelerated growth model proposed by Dorogovtsev {\it et al.} \cite{accel}. 
(x) The fitness model \cite{fitness} with a flat fitness distribution.  
(xi) The stochastic model for the protein interaction networks
introduced by Sol\'e {\it et al.} \cite{sole}. 
All those networks (i)-(xi) exhibit a power-law behavior 
in the BC distribution with the exponent $\eta \approx 2.2(1)$. 
Detailed properties of each network are listed in Table 1. 
The representative BC distributions for real world networks
(i), (iii), and (iv) are shown in Fig.1a.
 
The networks that we find to belong to the class II with $\eta=2.0$ 
include:
(xii) The Internet at the autonomous systems (AS) level as of October, 2001 \cite{internet2}. 
(xiii) The metabolic networks for 6 species of archaea in Ref.\cite{metabolic}.
(xiv) The WWW of {\tt nd.edu} \cite{www}. 
(xv) The BA model with $m=1$ \cite{ba}. 
(xvi) The deterministic model by Jung {\it et al.} \cite{jung}. 
In particular, the networks (xv) and (xvi) are of tree structure, 
where the edge BC distribution can be solved analytically. 
The detailed properties of each network are listed in Table 1.
The BC distributions for real world networks (xii) and (xiv) are shown in Fig.1b.
Since the BC exponents of each class are very close numerically, one may 
wonder if there exist really two different universal classes 
apart from error bar. To make this point clear, we plot the BC 
distributions for the BA model with $m=1,2$ and $3$ in Fig.2, 
obtained from large system size, $N=3\times10^5$. 
We can see clearly different behaviors between the two BC 
distributions for the cases of $m=1$ (class II) 
and of $m=2$ and $3$ (class I).\\

\noindent
{\sf
Topology of the Shortest Pathways\\}
To understand the generic topological features of the networks in each class, 
we particularly focus on the topology of the shortest pathways 
between two vertices separated by a distance $d$. 
Along the shortest pathways, 
we count the total number of vertices ${\cal M}(d)$ lying on 
these roads, averaged over all pairs 
of vertices separated by the same distance $d$. 
Adopting from the fractal theory, ${\cal M}(d)$ is called 
the ``mass-distance'' relation. We find that it behaves 
in different ways for each class; 
For the class I, ${\cal M}(d)$ behaves nonlinearly (Figs.3a-b), 
while for the class II, it is roughly linear (Fig.3c-d).

For the networks belonging to the class I such as the PIN2
(iii) and the metabolic network for eukaryotes (iv), 
${\cal M}(d)$ exhibits a non-monotonic behavior (Fig.3a-b), {\it viz.},
it exhibits a hump at $d_h\approx 10$ for (iii) or 
$d_h \approx 14$ for (iv). To understand why such a hump arises,
we visualize the topology of the shortest pathways between a pair of vertices,
taken from the metabolic network of a eukaryote organism, 
{\it Emericella nidulans} ({\it EN}), 
as a prototypical example for the class I. 
Fig.4a shows such a graph with linear size 26 edges ($d=26$),
where an edge between a substrate and an enzyme is taken as the unit of length.
From Fig.4a, one can see that there exists a blob structure 
inside which vertices are multiply connected, 
while vertices outside are singly connected. 
What is characteristic for the class I is that 
the blob is localized in a small region. To see this,
we measure the mass density $m(r;d)$, 
the average number of substrates or enzymes
located at position $r$ $\bigl(\sum_{r=1}^{d}m(r;d) = {\cal M}(d)\bigr)$. 
The average is taken over 
all possible pairs of vertices ($56$ pairs), separated 
by the same distance $d=26$. Note that the metabolic 
network is directed, so that the position $r$ is uniquely 
defined. 
As shown in Fig.5, we find that $m(r;d)$ is sharply peaked at $r=3$, 
and is larger than 1 only at $r=2,4,$ and 6 
for substrates.  
Thus the blob structure is present, even after taking averages, 
and is localized in a small 
region of size $d_b\simeq4$$\sim$$5$, centered at almost the same position 
$r\approx 3$$\sim$$4$ for different pairs of vertices. 
The blob size $d_b$ can be measured in another way. 
In a given graph of the shortest pathways, we delete 
singly connected substrates successively until none is left
and measure the linear size of the remaining structure. 
When averaged over all pairs of vertices with separations $d > 10$,
it comes out to be $d_b \approx 4.5$, 
well consistent with the value obtained previously for $d=26$ only. 
Due to this blob structure, the mass-distance relation 
increases abruptly across $d=4$ as shown in Fig.3b. 

Next, we measure the average mass of blob, that is, 
the number of vertices inside a blob for a given 
graph of the shortest pathways with separation $d$, averaged 
over all pairs of vertices with the same separation. 
We find that the average blob mass is broadly distributed 
in the range $3 < m_b < 23$. In particular, relatively heavy
blob masses, $m_b=15$$\sim$$23$, mainly comes from the graphs 
whose linear size is $d=8$$\sim$$14$.  
Due to those blobs with heavy mass, the mass-distance 
relation exhibits a hump, and decreases at around $d=14$$\sim$$16$, 
beyond which, the mass ${\cal M}(d)$ increases linearly 
by the presence of singly connected vertices. 
In short, the anomalous behavior in the mass-distance 
relation is due to the presence of a {\it compact and localized}
blob structure in the topology of the shortest pathways between 
a pair of vertices for the metabolic network of eukaryotes. 
We have checked the mass-distance relations and the graphs 
of the shortest pathways for other networks belonging to the class I 
such as the PIN2 and metabolic 
networks for other organisms, and found that such topological 
features are
generic, generating the anomalous behavior in the mass-distance 
relation. 
It still remains a challenge to derive the BC exponent 
$\eta \approx 2.2$ analytically from such structures.

For the class II, the mass depends on distance linearly, 
${\cal M}(d)\sim A d$ for large $d$ (Fig.3c-d). 
Despite the linear dependence, the shortest pathway topology for the 
case of $A>1$ is more complicated than that of the simple tree
structure where $A=1$. Therefore,
the SF networks in the class II are subdivided 
into two types, called the class IIa and IIb, respectively. For the class IIa, 
$A >1$ and the topology of the shortest pathways 
includes multiply connected vertices (Figs.4b and 4c), while for the class IIb, 
$A \sim 1$ and the shortest pathway is almost singly 
connected (Fig.4d). Examples in real world networks in the class 
IIa are the Internet at the AS level ($A\sim4.5$) and the metabolic network 
for archaea ($A\sim 2.0$), while that in the class IIb is the WWW ($A\sim1.0$).

Let us examine the topological features of the shortest pathways 
for the networks in the class IIa and IIb more closely. 
First, for the class IIa, we visualize in Fig.4b a shortest 
pathway in the Internet system between a pair of vertices separated by 10 
edges, the farthest separation. 
It contains a blob structure, but the blob is rather 
extended as $d_b=5$, comparable to the maximum separation $d=10$. We 
obtain $d_b=5$ for $d=11$ for another system. 
For comparison, $d_b\approx4.5$ for $d=26$ in the class I. 
Moreover, 
the feature-less mass-position dependence $m(r;d)$ we have found implies 
that while most blobs are located almost in the middle 
of the shortest pathways, which seems to be 
caused by the geometric effect, there are a finite number of 
blobs located at the verge of the shortest pathways. 
Note that $m(r;d) =m(d-r; d)$ since the Internet is undirected.
Owing to the extended structure and the scattered 
location of the blob, the mass-distance 
relation exhibits the linear behavior, ${\cal M}(d)\sim 
A d$ with $A \approx 4.5$. 
The extended blob structure is also observed in the 
metabolic network for archaea (Fig.4c). 
Since the network in this case is directed, 
the symmetry in $m(r;d)$ does not hold.
However, 
the blob structure extends to almost one half of maximum separation, 
and the shortest pathways are very diverse, 
so that their topological property such as the 
mass-distance relation ${\cal M}(d)$ is similar to that of the Internet.

The WWW is an example belonging to the class IIb. For this network, the 
mass-distance relation exhibits 
${\cal M}(d) \sim 1.0 d$, suggesting that the topology 
of the shortest pathway is almost singly connected, which 
is confirmed in Fig.4d. When a SF network is of  
tree structure, one can solve the distribution of BC running through each
edge analytically, and 
obtain the BC exponent to be $\eta=2$. A derivation of this exact result is
presented in the Appendix.\\

\noindent
{\sf
Comparison of the Resilience under Attack\\}
So far we have investigated the topological features of 
the shortest pathways of SF networks of each class. 
Then what would be distinct physical phenomena originated
from such different topological features? 
Associated with this question, we investigate a problem 
of the resilience of network under a malicious attack. 
It is known that SF networks are extremely vulnerable 
to the intentional attack to a few vertices with high degree,
while it is very robust to random failures \cite{attack,havlin}. 
To compare 
how vulnerable a network in each different class is 
under such attacks, we first construct a directed network 
whose numbers of vertices and edges, and the degree distribution are 
identical to  those of the WWW (xiv), but whose BC exponent
is $2.2$. It can be generated, for example, by following the 
stochastic rule introduced in the directed static model \cite{load}.
For both the WWW in real world and the artificial model network, 
we remove vertices in the descending order of BC successively.
As vertices 
are removed, both the mean distance $\langle d\rangle$ between two vertices, 
known as the diameter, and the relative size of the giant cluster 
$S$ are measured as a function of the fraction of removed vertices $f$.
As can be seen in Fig.6a, 
the diameter of the WWW with $\eta=2.0$ (class IIb) increases more rapidly 
than that with $\eta=2.2$ (class I) and shows 
discrete jumps while vertices are removed.
Also the relative size of the largest cluster decreases more rapidly 
for $\eta=2.0$ than for $\eta \approx 2.2$ (Fig.6b).   
This behavior arises from the fact that the shortest pathway 
consists of mainly singly connected vertices for the class IIb, 
so that there is no alternative pathways with the same distance 
when a single vertex lying on the shortest pathway is removed.  
For the Internet in real 
world with $\eta \approx 2.0$ in the class IIa and an artificial 
network with $\eta \approx 2.2$ with the 
same numbers of vertices and edges and the identical degree distribution,
the differences in the diameter $\langle d\rangle$ 
and in the relative size $S$ of the largest cluster
appear to be rather small (Figs.6c-d), in comparison
to the case of the WWW (Figs.6a-b). 
This is because the shortest pathways are multiply 
connected for the class IIa.\\

\noindent
{\sf
Conclusions\\}
In conclusion, we have found that the betweenness centrality can determine
the universal behavior of SF networks. By examining a variety of real 
world and artificial SF networks, we have observed two distinct universality 
classes whose BC exponents are $\eta\simeq 2.2(1)$ (class I) 
and $2.0$ (class II), respectively.
The mass-distance relation is introduced to characterize the topological
features of the shortest pathways. It shows a hump for the class I networks
due to compact and localized blobs in the shortest pathway topologies, while
it is roughly linear for the class II ones which are more or less tree-like.
The class II networks can further be divided into two types depending on
whether the shortest pathway topology contains diversified pathways (class
IIa) or mostly singly connected ones (class IIb). Distinct features 
of the resilience under attack arising from the different 
topologies of the shortest pathways are also identified. 
Since SF networks show the small world property, the topology of the 
shortest pathways should be of relevance for characterizing the network
geometry. Indeed the mass-distance relations for different universality
classes show different behaviors. Such a relation between the universality
class and the topological features of the shortest pathways may be
understood from the perspective of the fact that the geometric fractal
structure of the magnetic domains in equilibrium spin systems at criticality
can classify the universality classes.
Further characterizations
in static and dynamic properties and possible evolutionary origin of 
the universality classes are interesting questions left for future study.\\

\noindent
{\sf Acknowledgments\\}
{
This work was supported by the Korean Research Foundation 
(Grant No. 01-041-D00061) and BK21 program of MOE, Korea.\\}

\noindent
{\sf
APPENDIX\\} 
Here we present the analytic derivation of the BC distribution 
for a tree structure, however, the derivation is carried out
for the edge BC rather than the vertex version\footnote{See Szab\'o 
{\it et al.} (cond-mat/0203278, 2nd version) for a mean field treatment of 
the vertex BC problem on trees.}.
The edge betweenness centrality is 
defined on edges as in Eq.(1), with the subscript $k$ now denoting a bond. 
Without any rigorous 
proof, we assume that the distributions of vertex BC and 
edge BC behave in the same manner particularly on tree structures, 
which is confirmed by numerical simulations. We have also checked the identity
between the vertex BC and the edge BC distributions for a deterministic
model of scale-free tree introduced by Jung {\it et al.} \cite{jung}, which
will be published elsewhere.

We consider a growing tree network such as the BA type model with 
$m=1$, where a newly introduced vertex attaches an edge to an 
already existing vertex $j$ with the probability proportional to 
its degree as $(k_j+a)/\sum_{\ell} (k_{\ell}+a)$. 
Then the network consists of $N(t)=t+1$ vertices and $L(t)=t$~ edges 
at time $t$. 
The stationary degree distribution is of a power law with $\gamma=3+a$ \cite{bu,porto}.
Each edge of a tree divides the vertices into two groups attached to either
sides of the edge.
Let $P_s(m,t)$ be the probability 
that the edge born at time $s$ bridges a cluster with $m$ vertices on the 
descendant side and another with remaining $t+1-m$ vertices on the ancestor side. 
Due to the tree structure, the BC running through that edge born at $s$ 
is given as $g=2m(t+1-m)$, independent of the birth time $s$. 
The probability $P_s(m,t)$ evolves as a new vertex attaches 
to one of the two clusters. 
The rate equation for this process is written as 
\renewcommand{\theequation}{A\arabic{equation}}
\setcounter{equation}{0}
\begin{equation}
P_s(m,t+1)=r_1(m,t)P_s(m,t)+r_2(m-1,t)P_s(m-1,t), 
\label{rateeq}
\end{equation}
where $r_1(m,t)$ is the probability that a new vertex attaches 
to the cluster with $(t+1-m)$ vertices on the ancestor side, 
and $r_2(m-1,t)$ with $(m-1)$ vertices on the descendant side. 
They are given explicitly as  
\begin{equation}
r_1(m,t)=1-r_2(m,t)=\frac{(2+a)(t-m)+at+1}{2t+a(t+1)}.
\end{equation}
Since the amount of the BC on the edge $s$ is independent 
of the birth time, we introduce $P(m,t)$, 
\begin{equation}
P(m,t)={1 \over t}\sum_{s=1}^t P_s(m,t), 
\end{equation}
which is the probability for a certain edge to locate between 
two clusters with $m$ and $t+1-m$ vertices averaged over its birth time. 
The BC on that edge is still given by $2m(t+1-m)$.     
In terms of $P(m,t)$, Eq.(\ref{rateeq}) can be written as 
\begin{eqnarray}
(t+1)P(m,t+1) &=& r_1(m,t)tP(m,t)\cr 
& &+r_2(m-1,t)tP(m-1,t).
\label{rateeq1}
\end{eqnarray}
In the limit of $t\rightarrow \infty$, one may rewrite $P(m,t)$ 
in a scaling form, 
$P(m,t)={\cal P}(m/t)$ and then Eq.(\ref{rateeq1}) is rewritten as 
\begin{equation}
\nonumber
(t+1){\cal P}(x)-t{\cal P}(x)\simeq-x \frac{d{\cal P}(x)}{dx}-{\cal P}(x)
\end{equation}
where $x=m/t$ and the approximation ${\cal P}(x-1/t)\simeq 
{\cal P}(x)-(1/t){d{\cal P}(x)}/{dx}$ has been used. 
From this we obtain that 
\begin{equation}
{\cal P}(x)\sim \frac 1 {x^2},
\label{px}
\end{equation}
independent of the tuning parameter $a$.
Using $g=2(t+1-m)m \sim 2t^2 x$ for large $t$ and finite $m$, Eq.(\ref{px}) becomes 
\begin{equation}
P_B(g)\sim \frac{1}{g^2}.
\end{equation}
Thus $\eta=2$ is obtained for the tree structure, independent of $\gamma>2$.
General finite size scaling relations for $P_B(g)$ are discussed in Ref.\cite{kolkata}.\\

%
%

\vspace{1cm} 

\noindent{\sf Figure Legends\\}

\noindent
Fig. 1  {\sf The BC distributions of real world networks.}\\

\noindent
(a) Networks belonging to the class I: Co-authorship network ($\times$), 
Core of PIN of yeast by Ito {\it et al.} (+),
and metabolic network of {\it EN} ({\Large $\diamond$}).
The solid line is a fitted line with a slope $-2.2$.
(b) Networks belonging to the class II:
WWW of {\tt nd.edu} ({\Large$\circ$}) and Internet AS as of October, 2001 ($\square$).
The solid line has a slope $-2.0$.\\

\noindent
Fig. 2  {\sf Comparison of the BC distributions for the two classes.}\\

\noindent
BA model with $m=1$, 2, and 3 are simulated for large system size, 
$N= 3\times 10^5$, averaged over 10 configurations.
The dotted line has a slope $-2.0$ and the dashed one, $-2.2$.\\

\noindent
Fig. 3  {\sf  The mass-distance relation $\cal M$$(d)$.}\\

\noindent
(a) Core of PIN of yeast obtained by Ito {\it et al.}.
(b) Metabolic networks of eukaryotes. Data are averaged over all 
5 organisms in Ref.\cite{metabolic}. Note that in this case we count 
only substrates for ${\cal M}(d)$.
(c) Internet AS as of October, 2001.
(d) WWW of {\tt nd.edu}.\\

\noindent
Fig. 4  {\sf Topology of the shortest pathways.}\\

\noindent
(a) The metabolic network of {\it EN} (eukaryote) of length 26.
(b) The Internet AS of length 10.
(c) The metabolic network of {\it Methanococcus jannaschii} (archae) of length 20.
(d) WWW of {\tt nd.edu} of length 20.
In (a) and (c), circles denote substrates and rectangles denote intermediate states.\\

\noindent
Fig. 5  {\sf The mass density.}\\

\noindent
$m(r;d)$ for {\it EN} with $d=26$. Circles denote substrates and rectangles intermediate states.\\

\noindent
Fig. 6  {\sf Attack vulnerability of the scale-free networks.}\\

\noindent
The WWW ($\eta=2.0$) ($\blacksquare$) and the artificial directed SF network with $\eta=2.2$ ({$\square$}), the Internet ($\eta=2.0$) ({\Large$\bullet$}) and the artificial undirected SF network with $\eta=2.2$ ({\Large$\circ$}): Changes in network diameter (a, c) and the relative size of the largest cluster (b, d) are shown as a function of $f$, the fraction of removed vertices measured in percent (\%).\\

\noindent{\sf Table Legend\\}

\noindent
Table 1  {\sf Natures of diverse SF networks.}\\

\noindent
Tabulated for each network are the size $N$, 
the mean degree $\langle k\rangle$, 
the degree exponent $\gamma$, 
and the betweenness centrality exponent $\eta$.\\

\begin{figure}[!h]
\centerline{\epsfxsize=8cm \epsfbox{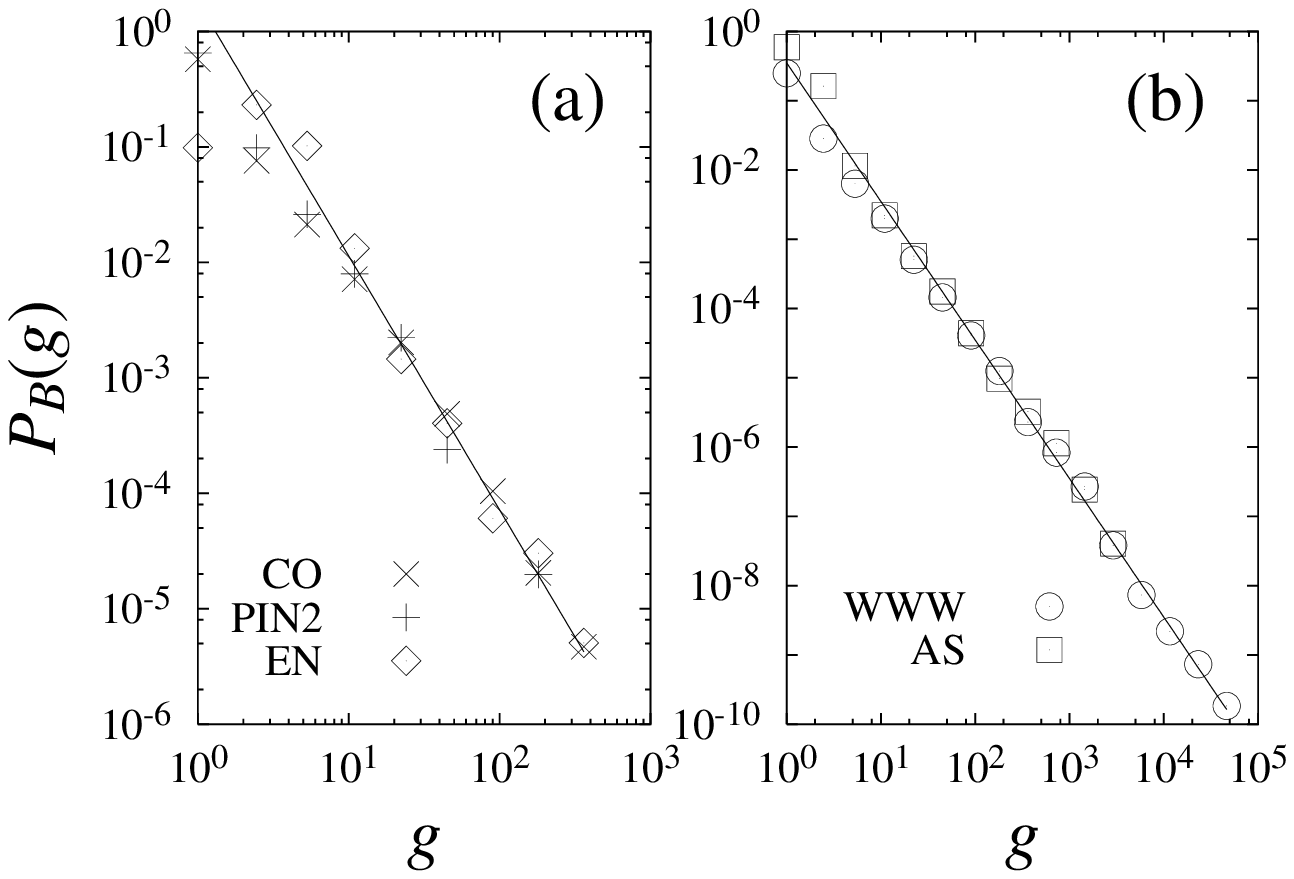}}
\caption{ Goh et al. }
\end{figure}

\begin{figure}[!h]
\centerline{\epsfxsize=8cm \epsfbox{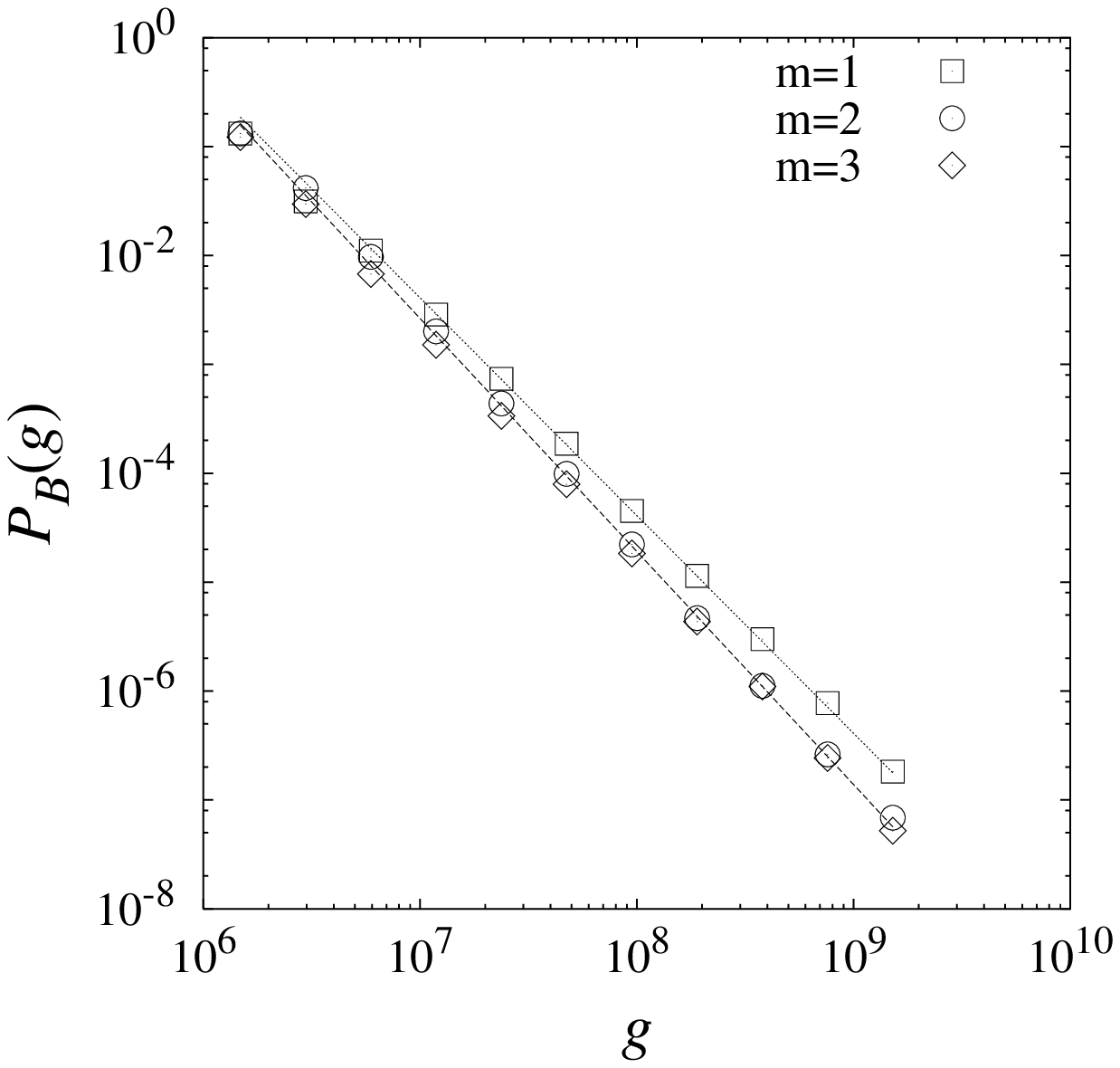}}
\caption{Goh et al.}
\end{figure}

\begin{figure}[!h]
\centerline{\epsfxsize=8cm \epsfbox{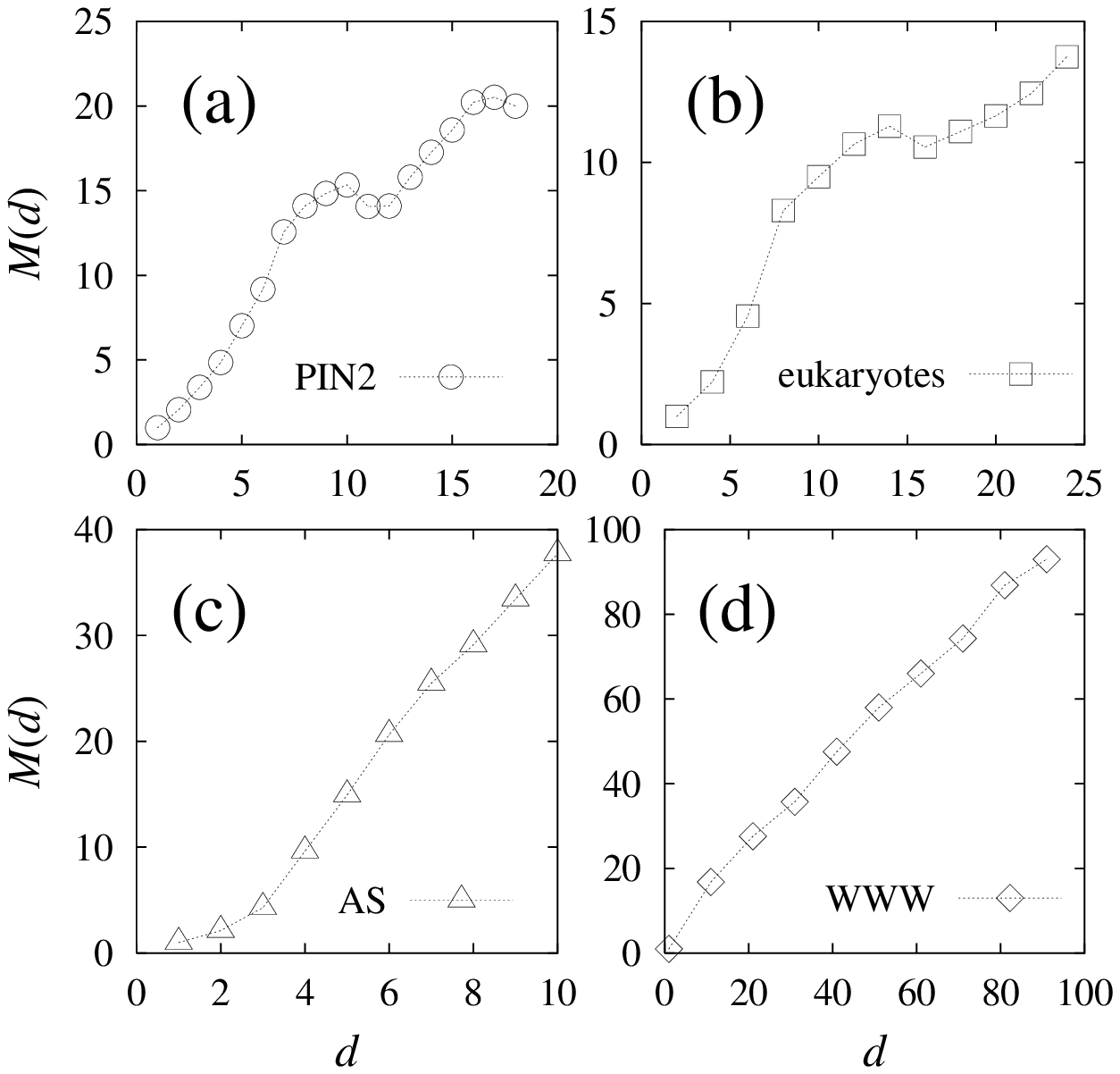}}
\caption{Goh et al.
}
\end{figure}

\begin{figure}[!h]
\centerline{\epsfxsize=8cm \epsfbox{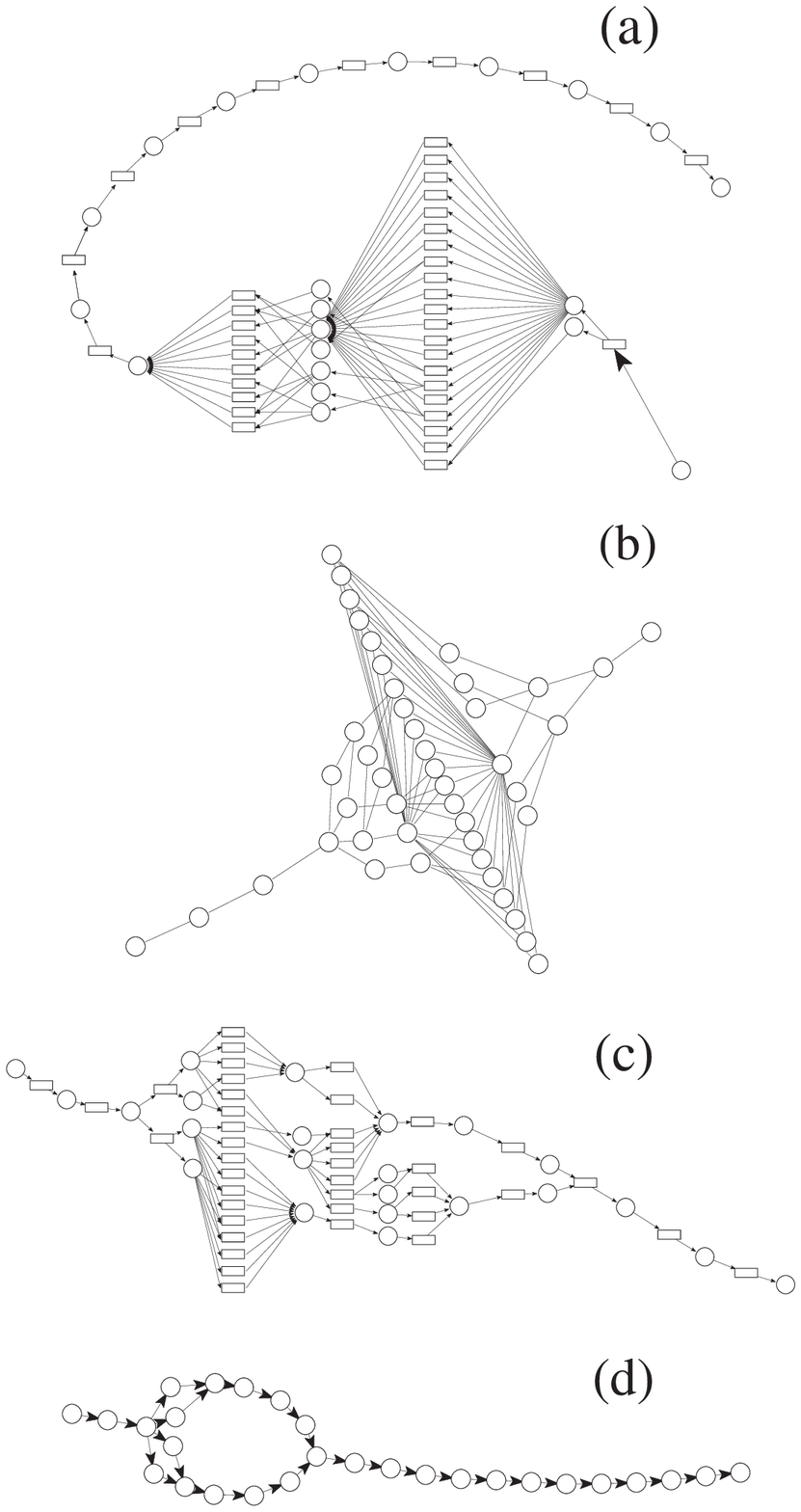}}
\caption{Goh et al.}
\end{figure}

\begin{figure}[!h]
\centerline{\epsfxsize=8cm \epsfbox{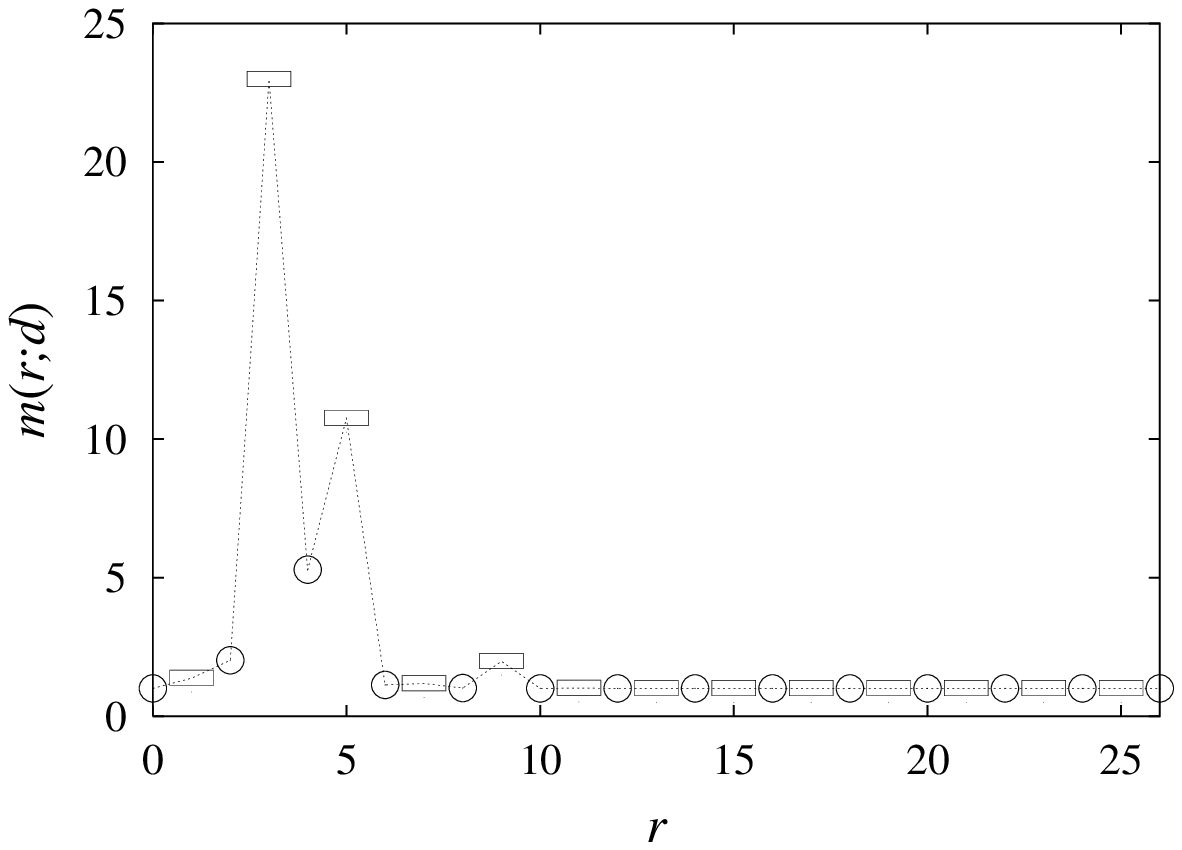}}
\caption{Goh et al.}
\end{figure}

\begin{figure}[!h]
\centerline{\epsfxsize=8cm \epsfbox{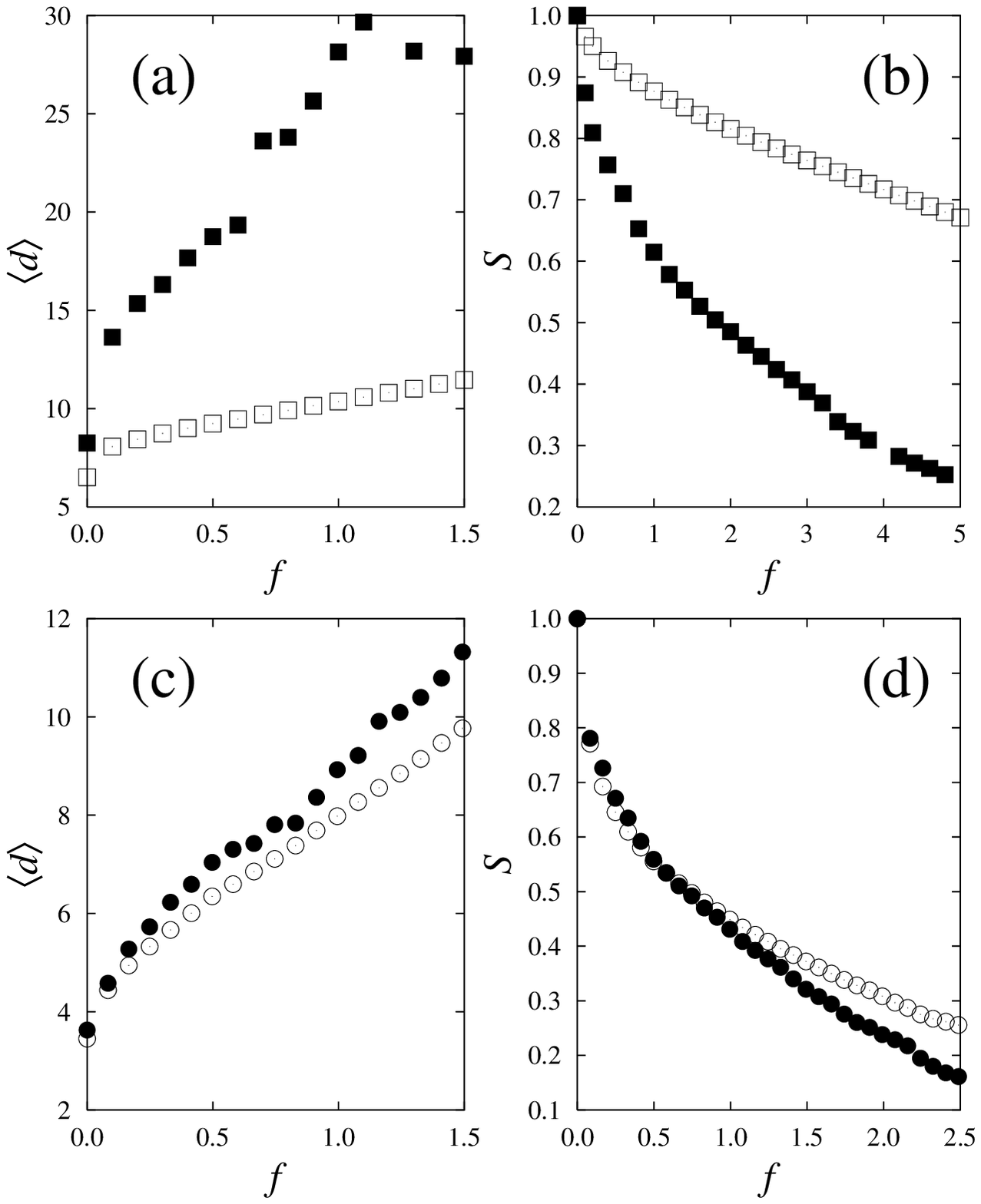}}
\caption{Goh et al.}
\end{figure}

\begin{table*}
\caption{Goh et al.}
\small
\hspace{-1.5cm}
\begin{tabular}{c|rlccccc}
\hline\hline
Class & &System & $N$ & $\langle k\rangle$ & $\gamma$ & $\eta$ & Ref. \\
\hline
 &(i)& Co-authorship & 205 202 & 11.8 & 2.2(1) & 2.2(1) & \cite{coauthor}\\
 &(ii)& PIN1 & 1846 & 2.39 & 2.4 (exp.\ cut-off) & 2.2(2) &\cite{pin}\\
 &(iii)& PIN2 & 797 & 1.96 & 2.7(1) & 2.2(1) & \cite{ito}\\
 &(iv)& Metabolic (eukaryotes, bacteria)& $\sim$$10^3$ & 2-4 & 2.0-2.4 & 2.2(1) & \cite{metabolic}\\
Class I &(v)& BA model ($m\ge2$) & $3\times10^5$ & 2$m$ & 2.0-3.0 &2.2(1) &  \cite{ba}\\
 &(vi)& HA model & $10^5$ & ${\cal O}(1)$ & 3.0(1) & 2.2(1)  & \cite{ha}\\
 &(vii)& Copying model & $10^4$ & 4 & 2.0-3.0 & 2.2(1) & \cite{copying}\\
 &(viii)& Static model & $10^4$ & 4,6,8 & 2.0-3.0 & 2.2(1) & \cite{load}\\
 &(ix)& Accelerated growth model & $10^4$ & ${\cal O}$(1) & 3.0(1) &2.2(1)  & \cite{accel}\\
 &(x)& Fitness model & $10^4$ & 4 & 2.25 &2.2(1) &  \cite{fitness}\\
 &(xi)& PIN model & $10^4$ & $\sim$2 & 2-3 &2.2(1)  & \cite{sole}\\
\hline
Class IIa &(xii)& Internet AS & 12 058 & 4.16 & 2.2 &  2.0(1) & \cite{internet2}\\
 &(xiii)& Metabolic (archaea) & $\sim$$10^3$ & 2-4 & 2.0-2.3 &  2.0(1) & \cite{metabolic}\\
\hline
 &(xiv)& WWW & 325 729 & 4.51 & 2.1/2.45 & 2.0 &  \cite{www}\\
Class IIb &(xv)& BA tree ($m=1$) & $\infty$ & 2 & $>2.0$ &  2.0 & \cite{ba}\\
 &(xvi)& Deterministic tree & $\infty$ & 2 & $>2.0$ &  2.0 & \cite{jung}\\
\hline\hline
\end{tabular}
\end{table*}

\end{document}